\documentclass[a4paper,onecolumn,11pt]{quantumarticle}
\pdfoutput=1
\usepackage{authblk}
\usepackage[utf8]{inputenc}
\usepackage[english]{babel}
\usepackage[T1]{fontenc}
\usepackage{amsmath}
\usepackage{bbold}
\usepackage{dsfont}
\usepackage{graphicx}
\usepackage{dcolumn}
\usepackage{bm}
\usepackage[colorlinks, allcolors={blue}]{hyperref}
\usepackage{braket}
\usepackage{subcaption}
\usepackage{relsize}
\usepackage{amsthm}
\newtheorem{lemma}{Lemma}
\usepackage{mathtools}
\usepackage{amsmath,amssymb,amsthm}
\usepackage{enumitem}
\newtheorem{theorem}{Theorem}
\newtheorem{definition}{Definition}
\usepackage{comment}
\excludecomment{ignore}  
\newtheorem{problem}{Problem}

\begin{document}

\title{Quantum Circuit Cutting: Complexity and Optimization}
 
\newcommand{\affBIU}{Faculty of Engineering and the Institute of Nanotechnology and Advanced Materials, Bar-Ilan University, Ramat Gan 5290002, Israel}
\newcommand{\affIQS}{Institute for Quantum Studies, Chapman University, Orange, California 92866, USA}
\newcommand{\affNVIDIA}{NVIDIA, Hakidma 26, Ofer Industrial Park, Yokneam 2069203, Israel}
\newcommand{\affTECHNION}{The Henry and Marilyn Taub Faculty of Computer Science, Technion, Israel.}

\author{Yuval Idan}
\orcid{0009-0009-9632-3229}
\affiliation{\affBIU}
\affiliation{\affNVIDIA}

\author{Eitan Zahavi}
\affiliation{\affNVIDIA}

\author{Elad Mentovich}
\affiliation{\affNVIDIA}

\author{Eliahu Cohen}
\affiliation{\affBIU}\affiliation{\affIQS}

\author{Shmuel Zaks}
\affiliation{\affNVIDIA}
\affiliation{\affTECHNION}

\maketitle

\begin{abstract}
The current noisy intermediate-scale quantum (NISQ) era is characterized by substantial errors and noise, which limit the practical feasibility of deep, many-qubit circuits. To address these constraints, quantum circuit cutting has emerged as a promising tool. Recently, there has been significant research on methods for performing such cutting effectively. In this work, the duality between quantum circuits and classical graphs---specifically, directed acyclic graphs (dags)---is leveraged to analyze the complexity of finding an optimal circuit-cutting configuration that minimizes the number of cuts. After developing a rigorous graph-theoretic framework, the complexity of identifying cut locations that partition a given quantum circuit into smaller fragments is characterized. The corresponding graph-combinatorial task is then defined, and the resulting partition problem is shown to be NP-complete. Furthermore, even a simplified version of the problem, restricted to circuits composed only of one- and two-qubit gates, is shown to be NP-complete. Finally, based on these constraints, an algorithm grounded in satisfiability modulo theories (SMT) is proposed to find optimal cuts when the number of qubits per partition is bounded. This work therefore provides a complexity-theoretic characterization of cut placement and a practical solver for bounded-size decompositions.
\end{abstract}

\section{Introduction}
In the noisy intermediate-scale quantum (NISQ) era \cite{preskill2018quantum}, scaling up the number of qubits in a single coherent device remains a formidable challenge. This challenge hinders the realization of quantum advantage that is expected from fault-tolerant quantum computers. There has been significant progress in various hardware platforms, from trapped ions to superconducting circuits \cite{fukui2022building,bruzewicz2019trapped}, yet the exponential growth of the state space imposes severe resource constraints \cite{kim2023evidence, bharti2022noisy}, necessitating the use of high-performance computing (HPC) to implement larger-scale computations, utilizing various protocols which optimize the computation.
One method aimed at enabling quantum computation over large Hilbert spaces in the NISQ era is the circuit-cutting protocol (also known as circuit knitting). It allows quantum programs that require more qubits than are currently available to be distributed across multiple smaller quantum processing units (QPUs), aided by classical HPC. The main drawback is an exponential computational overhead \cite{cutting_Haram}.
In \cite{BravyiSmithSmolin2016}, it was proved that any sparse quantum circuit on \(n{+}k\) qubits can be simulated by a sparse circuit on \(n\) qubits with classical post-processing overhead of \(2^{O(k)}\mathrm{poly}(n)\). Later, in \cite{HarrowLowe2025} a cluster-based simulation scheme that uses edge (wire) cuts via Pauli-basis decompositions to run large circuits on multiple smaller processors. The run time of such processes increases exponentially with the number of cuts.
These methods build on the quasi-probabilistic local operations and classical communication (LOCC) framework~\cite{PiveteauSutter2024} and can be further optimized by reducing the sampling overhead by using the suitable decompositions.
Subsequent works \cite{SchmittPiveteau2025,Piveteau2025} extend this framework to clusters of two‑qubit unitaries and to instrument‑based classical side information, providing semidefinite‑programming lower bounds on the sampling cost under LOCC constraints.
Within this framework, \cite{HarrowLowe2025} formalized two types of cuts: spacelike and timelike. In this work, we focus exclusively on timelike cuts, which correspond to cutting qubits. This focus on timelike cutting is motivated by the fact that the sampling overhead associated with spacelike cutting depends on the specific unitary being cut and is quantified by the so called entanglement robustness, $\gamma_S(U)$ \cite{PiveteauSutter2024, SchmittPiveteau2025,Piveteau2025} (whose calculation is non-trivial), rather than by a uniform overhead given by a constant factor per cut (like the one we use here). This work assumes an optimal circuit knitting protocol (in terms of spectral decompositions) and directs the attention to the optimal placement of qubit cuts. Restricting cuts to single qubits avoids the substantial overhead associated with cutting multi-qubit operators \cite{ufrecht2023cutting}.
The main framework in our analysis is  converting quantum circuits into tensor networks \cite{biamonte2017tensor}, which transform the unitary evolution into an extended graphical formalism that is amenable to graph-theoretic analysis, which allows to  use rigorous mathematical theorems from graph theory.

This work characterizes the computational complexity of optimizing timelike (wire) cuts for an arbitrary quantum circuit.
Using a circuit--graph correspondence, we represent the circuit as a legal directed acyclic graph (dag) and model a wire cut by an \emph{edge duplication} operation, which splits the wire into two separate wires.
This terminology isolates the combinatorial core of circuit cutting: the number of duplications determines the classical post-processing cost, so the central objective is to minimize duplications while satisfying bounded cluster-size constraints. Our objective is to determine whether the resulting legal dag can be partitioned into at most $\alpha$ clusters, each containing at most $k$ input/output wires, using at most $\beta$ edge duplications. We refer to this task as the graph duplication (GD) problem.
Equivalently, one may consider the optimization variant that minimizes the number of duplications required to satisfy these cluster-size constraints.
We divide quantum cutting protocols into classes according to $\beta$, which denotes the number of duplications required to satisfy our constraints. In particular, two cases are considered: $\beta=0$ and $\beta>0$. The case $\beta=0$ refers to circuits that can already be carried out by multiple local computers without any edge duplication, whereas $\beta>0$ requires duplicating at least one edge.

Our main results are the following. For a constant number of connected components and for any constant  $\beta$, GD can be decided in polynomial time. In contrast, GD becomes NP-complete when the number of components is not bounded or the number of duplications (cuts) $\beta$ is not bounded.
In particular, NP-completeness holds already for $\beta=0$ on forests, and persists under additional restrictions such as 2-legal dags (corresponding to circuits composed only of one- and two-qubit gates which can be compiled into a universal gate set \cite{1995elementary}). 
Finally, a formulation of satisfiability modulo theories (SMT) \cite{brandhofer2023optimal}, compatible with our definitions and constraints, is designed and used to construct a solver that can be applied to general quantum circuits.

The manuscript is structured as follows. Sec. \ref{Sec: quantum_circ_tensor} establishes connections between the quantum-circuit model, tensor networks, and dags, building on \cite{cutting_Haram}. Sec. \ref{Sec: circuit_cutting} briefly presents the circuit-cutting protocol and its utility. Sec. \ref{sec:Basic definitons} outlines graph-theoretic definitions and constraints used throughout the work. After defining the graph under study—namely, the legal dag—Sec. \ref{sec:operations on dag} introduces operations on this constructed graph. Sec. \ref{sec:graph duplication} then introduces the graph duplication (GD) problem. Sec.  \ref{Sec:resutls} presents the main results where complexity of the GD problem is analyzed across various test cases, and Sec. \ref{Sec:smt_solver} establishes an SMT solver that provides solutions to the circuit-cutting problem. Finally, Sec. \ref{Sec: DISCUSSION} concludes with a discussion of the obtained results and various directions for future research.


\section{Quantum circuit model and tensor networks}\label{Sec: quantum_circ_tensor}
Tensor networks are a useful tool to represent physical systems in a mathematically elegant and compact way \cite{biamonte2017tensor,selby2017process,biamonte2019lectures, bridgeman2017hand}. According to \cite{cutting_Haram}, every quantum simulation $C$ with a measurement outcome function $f:\{0, 1\}^n \rightarrow [-1, 1]$ can be represented as a pair $(C, f)$, which can be transformed into a tensor network $(G, \mathcal{A})$. Here, $G = (V, E)$ is a  dag and the initial state of each qubit is directed into a measurement device, $\mathcal{A}$ is a collection of associated tensors, which are unitary operators (denoted as a box: $\Box$) in the quantum algorithm. The rank of each tensor is equal to the number of indices, one for each  input and output Hilbert spaces. In general, the tensors, corresponding to quantum maps, can be unitary (hence the number of indices that enter the tensor and the number of indices that leave the tensor are equal) or non-unitary (the numbers may differ). In this work, the only two non-unitary tensors are qubit preparation (denoted as $\triangleleft$), and measurement (denoted as $\triangleright$).

For example, the initial state of the quantum circuit  $\mathcal{C}$ can be $n$  qubits  encoded (without loss of generality) in the $|0\rangle$ state: 
\begin{equation}\label{initial_state}
\ket{\psi_{\rm in}} = \ket{0}^{\otimes n}, 
\qquad 
\psi_{\rm in}(i_0,\ldots,i_{n-1}) := \braket{i_0\ldots i_{n-1}|\psi_{\rm in}}.
\end{equation}
Then, a unitary transformation that operates on the aforementioned tensor network has a $2n$ rank, because each input Hilbert space is being unitarily transformed ($\hat{U} : \mathcal{H}_{in} \rightarrow \mathcal{H}_{out}$), such that $dim(\mathcal{H}_{in}) = dim(\mathcal{H}_{out})$, i.e. the dimensions of the Hilbert spaces are equal. More intuitively,  qubits do not ``form'' or ``disappear'' during the computation. Continuing with the example, the notation for such a process is given as
\begin{equation} \label{unitary_tesnor}
\hat U 
= \sum_{\mathbf i,\mathbf j \in \{0,1\}^n} U_{\mathbf j,\mathbf i}\,\ket{\mathbf j}\!\bra{\mathbf i}
= \sum_{i_0,\ldots,i_{n-1}}\sum_{j_0,\ldots,j_{n-1}}
U^{j_0\ldots j_{n-1}}_{i_0\ldots i_{n-1}}
\ket{j_0\ldots j_{n-1}}\!\bra{i_0\ldots i_{n-1}},
\end{equation}
where $\mathbf i=\left(i_0\ldots i_{n-1}\right)$, $\mathbf j=\left(j_0\ldots j_{n-1}\right)$.
As earlier mentioned, for a unitary evolution (assuming each edge is a qubit) the number of the $i$ indices is equal to the number of the $j$ indices. The state in Eq.~(\ref{initial_state}) evolves according to the unitary transformation in Eq.~(\ref{unitary_tesnor}) as follows:
\begin{equation}
\ket{\phi} = \hat{U} \ket{\psi_{\rm in}},
\qquad
\phi(\mathbf j) 
= \sum_{\mathbf i\in\{0,1\}^n} U_{\mathbf j,\mathbf i}\,\psi_{\rm in}(\mathbf i).
\end{equation}
A measurement device is a one-way tensor, whereby acting on the state $\ket{\phi}$, the result of the measurement is a scalar:
\begin{equation}
\langle M\rangle = \bra{\phi} M \ket{\phi}
= \mathrm{Tr}(M\,\rho_{\rm out}),
\qquad
\rho_{\rm out} := \ket{\phi}\!\bra{\phi}.
\end{equation}
The entire network is depicted in Fig.~\ref{Fig:tensor_evolution}.

\begin{figure}
    \centering
    \includegraphics[width=0.6\linewidth]{tensor_networks.png}
    \caption{Evolution of a quantum state within a tensor network. The initial state tensor is enclosed by the blue dashed line. The red dashed line highlights the output state resulting from the action of the unitary operator $\hat{U}$ on the initial state. Finally, the green dashed line encompasses the entire process, yielding a scalar value.
 }
    \label{Fig:tensor_evolution}
\end{figure}
When each qubit is measured individually, the measurement device \( M \) is modeled as a tensor product of local measurement devices with two outcomes, as described in~\cite{cutting_Haram}, for some local outcomes function \( f \).
This method bears some relation with \cite{huang2020predicting}, which also discusses the relation of local channels relative to a global quantum state.

Having established the structure of tensor networks, we now turn to dags. A dag is denoted by \( G = (V, E) \), where each edge \( e \in E \) represents the trajectory of a single qubit between two operations (tensors), and each vertex corresponds to a quantum operation acting on one or more qubits (tensor).
We denote the sets of input and output vertices by the sets \( In(G) \) and \( Out(G) \), respectively; these are the only tensors that are asymmetric (in the degree), while in all the others the number of incident edges is equal to the number of outgoing edges, preserving unitary evolution within the graph.
This work focuses on the general tensor-network architecture. 
Within this framework, the labeling of tensor legs may be permuted by inserting SWAP gates that exchange qubit indices. 
Since SWAP operations can be embedded within multi-qubit unitaries, the ordering of the output indices may not coincide with the ordering of the input indices. The SWAP gate may be inserted to perform cuts.
In the proposed framework, sequential single‑qubit gates are combined into a single tensor (vertex), since placing a cut between two consecutive single‑qubit operations does not yield a meaningful decomposition (do not change the number of qubits in each circuit). More generally, any sequence of rank‑\(d\) tensors in which the output index of one tensor directly feeds into the input index of the next is consolidated into a single rank‑\(d\) tensor (because we do not consider space-like cuts \cite{HarrowLowe2025}). These requirements preserve the overall quantum mapping while simplifying the tensor network by eliminating trivial cuts that do not contribute to computational considerations in our analysis.  

\section{Circuit cutting}\label{Sec: circuit_cutting}
\begin{figure}[ht]
    \centering
    \begin{subfigure}[b]{0.49\textwidth}
        \centering
        \includegraphics[width=\linewidth, height=4.9cm]{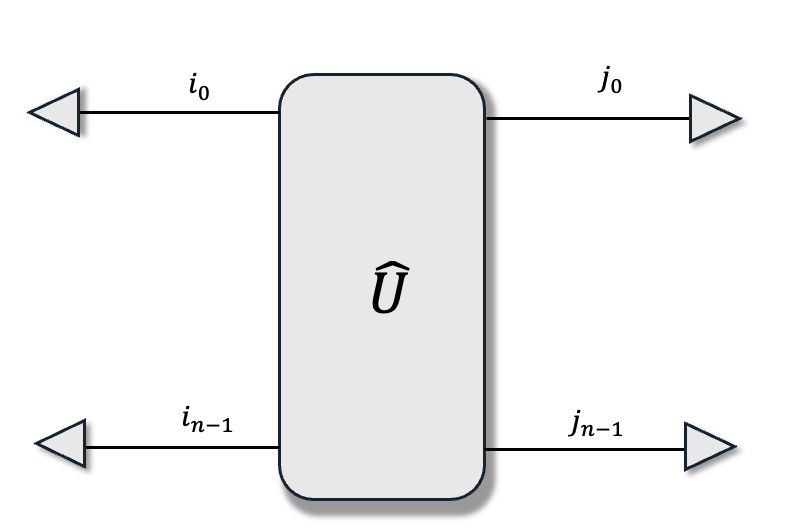}
        \caption{}
        \label{Fig:unitary}
    \end{subfigure}
    \hfill
    \begin{subfigure}[b]{0.49\textwidth}
        \centering
        \includegraphics[width=\linewidth]{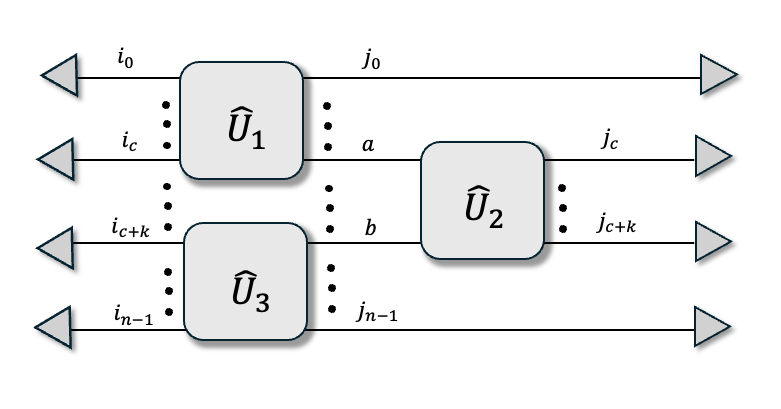}
        \caption{}
        \label{Fig:unitary_decomposition.png}
    \end{subfigure}
        \hfill
    \begin{subfigure}[b]{0.49\textwidth}
        \centering
        \includegraphics[width=\linewidth]{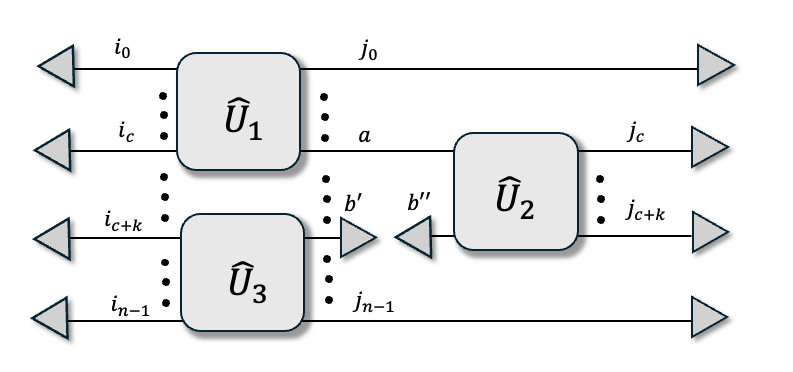}
        \caption{}
        \label{Fig:seperated}
    \end{subfigure}
    \hfill
    \begin{subfigure}[b]{0.49\textwidth}
        \centering
        \includegraphics[width=\linewidth]{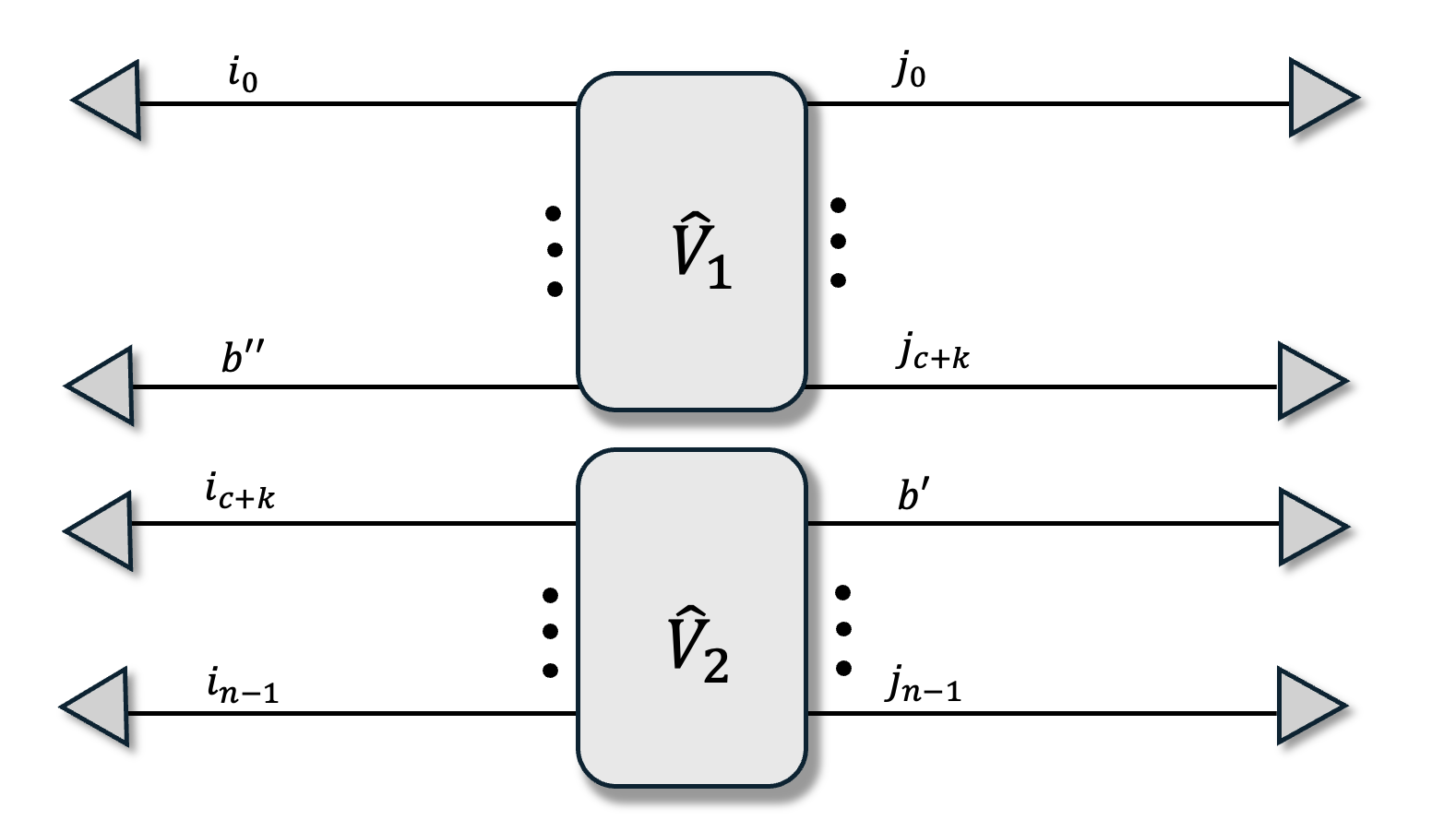}
        \caption{}
        \label{Fig:twou}
    \end{subfigure}
    \caption{(a) Rank-$2n$ unitary, which is not separable, and acting on all the qubits simultaneously. (b) Decompositions of  the previous  rank-$2n$ tensor $\hat{U}$ into lower-rank tensors. (c) Cutting the  unitary $\hat{U}$ into two sets of unitaries. (d) The contraction from Eq.~(\ref{eq:productten}). }
    \label{Fig:seperate}
\end{figure}

In the previous section, we introduced the relation between the circuit model, tensor networks and dags. The circuit-cutting method involves cutting an edge $e$ (Hilbert space $\mathcal{H}_{e}$) and replacing it with a measurement device and a prepared qubit. The measured and prepared qubits are correlated through projection onto, and preparation in, the same eigenstate,  creating two new edges $e'$ and $e''$ ($\mathcal{H}_e \rightarrow \mathcal{H}_{e'}\otimes \mathcal{H}_{e''}$) and two new vertices, one for each partition. One vertex is post-selected onto a quantum state (which is an eigenstate of one of the Pauli group operators), while the second qubit is pre-selected in the same eigenstate, establishing a quasi-causal order between the circuits. 
The fragments or partitions of the original graph are called clusters. Fig.~\ref{Fig:seperate} illustrates the process of generating two clusters from a single quantum circuit. The maximum number of clusters is at most $n$, which is the number of qubits in the original circuits.
Moreover, although it is possible to cut operators, we do not consider it to maintain a more straightforward and focused presentation. In addition, cutting qubits introduces relatively lower computational overhead (the number of samples required scales as $O(2^{4k}/\epsilon^2)$ \cite{cutting_Haram}, 
where $k$ is the number of cut edges), whereas cutting operators incurs a cost that depends on the number of input qubits and the degree of induced entanglement \cite{Piveteau2025,PiveteauSutter2024}, thereby significantly complicating the analysis and leading to a high sampling rate. Hence, already in Sec.~\ref{Sec: quantum_circ_tensor}, cutting vertices was excluded, i.e. we only allowed edge cuts. In addition, our constraints do not allow cutting the entire circuit horizontally, i.e., cutting all the qubits at once, or in a feedforward way. At least one qubit from the original quantum circuit must be connected in each sub-circuit. Let us examine an example of distributing a rank-$2n$ unitary 
(Eq.~(\ref{unitary_tesnor})) into smaller-rank tensors, thereby 
implementing the circuit-cutting protocol. 
Although $\hat{U}$ is written as a single global operator, it admits 
a decomposition into a sequence of local unitaries, forming a 
matrix product operator (MPO) representation. 
Such tensor can be written as a tensor network of three tensors:
\begin{equation}
U^{j_0\ldots j_{n-1}}_{i_0\ldots i_{n-1}}
= 
\left(U_1\right)^{j_0\ldots j_{c-1} a}_{i_0\ldots i_c}
\,
\left(U_2\right)^{j_c\ldots j_{c+k}}_{a\,\cdots\, b}
\,
\left(U_3\right)^{b\, j_{c+k+1}\ldots j_{n-1}}_{i_{c+k}\ldots i_{n-1}} .
\end{equation}
The unitary evolution described by \( \hat{U} \) is now expressed as a composition of three connected tensors, \( \hat{U}_1 \), \( \hat{U}_2 \), and \( \hat{U}_3 \) (see Fig.~\ref{Fig:unitary_decomposition.png}), each of which has a lower rank than the original tensor \( \hat{U} \). In this example, the most straightforward cuts correspond to the indices \( a \) or \( b \), under the assumption that the goal is to partition the network into two balanced clusters that together perform the overall unitary transformation.

Following \cite{cutting_Haram}, a single-qubit wire cut can be expressed using the
8-term decomposition:
\begin{equation}
A = \sum_{i=1}^{8} c_i\,\Phi_i(A),
\qquad
\Phi_i(A) := \mathrm{Tr}(A\,O_i)\,\rho_i,
\end{equation}
where $O_i\in\{I,X,Y,Z\}$ are Pauli observables, $\rho_i$ are the corresponding eigenprojectors,
and $c_i\in\{\pm\tfrac12\}$ (explicitly listed in \cite{cutting_Haram}).
Applying this identity to a tensor-network edge $e$ yields
\begin{equation}
T(G,A) = \sum_{i=1}^{8} c_i\, T(G_e, A^{(i)}),
\end{equation}
where $G_e$ is obtained by cutting $e$ and adding an observable vertex labeled by $O_i$ on the
source side and a state-preparation vertex labeled by $\rho_i$ on the sink side, and $A^{(i)}$
denotes the modified tensor assignment. For each edge cut, there are $8$ coefficients, one corresponding to each estimated projector. 
For $K$ edge cuts, the total number of coefficients scales exponentially with $K$, namely
\begin{equation}
|c| = 8^K.
\end{equation}
 
In general, it is possible to find a matrix product state (MPS) decomposition \cite{biamonte2019lectures} for any tensor network; however, such a decomposition may be computationally expensive, as it typically involves computing singular value decompositions (SVDs).

Now, assume that edge $b$ has been cut into two components $b' \  \text{and} \ b''$. Hence, one can use the well-known relations: $I = \sum_b\ket{b}\bra{b}$ and  $ \sum_{b,b',b''}\ket{b''}\braket{b''|b}\braket{b|b'}\bra{b'} =\delta_{b''}^{b'}I$. Inserting them into $\hat{U}$, leads to:
\begin{equation}
\hat{U}
=
\sum_{b',b''}
\left(U_1 U_2\right)^{\, j_0 \ldots j_{c+k}}_{\, i_0 \ldots i_c \, b''}
\,
\delta_{b''}^{\, b'}
\,
\left(U_3\right)^{\, b' \, j_{c+k+1} \ldots j_{n-1}}_{\, i_{c+k} \ldots i_{n-1}} .
\end{equation}
Now define
\begin{align}
V_1^{(b)} 
&:= 
\left(U_1 U_2\right)^{\, j_0 \ldots j_{c+k}}_{\, i_0 \ldots i_c \, b}, \\
V_2^{(b)} 
&:= 
\left(U_3\right)^{\, b \, j_{c+k+1} \ldots j_{n-1}}_{\, i_{c+k} \ldots i_{n-1}} .
\end{align}
Then, $\hat{U}$ can be written as a sum of tensor products of $V_1^{(b)}$ and  $V_2^{(b)}:$
\begin{equation}\label{eq:productten}
\hat{U}
=
\sum_{b}
V_1^{(b)} \otimes V_2^{(b)} .
\end{equation}
The unitary is thus decomposed into two tensor blocks linked by the bond index $b$, which encodes the shared Hilbert space. Each block may be simulated separately, and the contraction over 
$b$ is recovered via classical post-processing (and classical communication). The resulting circuit is illustrated in Fig.~\ref{Fig:twou}.

\section{Basic definitions}\label{sec:Basic definitons}
In this section we outline fundamental graph-theoretic definitions relevant to our analysis and later proofs.

\begin{definition}\label{definition:dag}
\em
A directed graph $G(V,E)$ consists of a (finite) set of vertices $V$ and a set of edges $E \subseteq V \times V$; that is, $e$ is an edge if $e=(a,b)$ for some $a,b \in V$. For an edge  $e=(a,b)$ we will assume that $a \neq b$.
\end{definition}
\begin{definition}
\em A directed graph is termed $acyclic$ if it does not contain directed cycles; such a graph is termed $dag$.  
\end{definition}
\begin{definition}
    
Given a directed graph \(G=(V,E)\) and a vertex \(v\in V\), the incoming and outgoing degrees of $v$ are
\[
  d_{\mathrm{in}}(v)
    = \bigl|\{\,u \mid (u,v)\in E\,\}\bigr|,
  \quad
  d_{\mathrm{out}}(v)
    = \bigl|\{\,u \mid (v,u)\in E\,\}\bigr|.
\]
\end{definition}
\noindent
In terms of quantum channels, each edge is a single qubit unitary evolution, and each vertex is multi-qubit unitary gate, acting on all incident qubit states. 

\begin{definition}
\em An $input ~vertex$ is a vertex $v \in V$ for which $d_{in}(v) = 0$. $In(G)$ denotes the set of input vertices; that is,  $In(G) = \{v ~|~ d_{in} (v) = 0 \}$.  
An $output ~vertex$ is a vertex $v \in V$ for which $d_{out}(v) = 0$. $Out(G)$ is the set of output vertices; that is,  $Out(G) = \{v ~|~ d_{out} (v) = 0\}$.
\end{definition}

Because the system is closed and the entire computation is unitary the number of qubits does not change through the algorithm. Therefore, for every vertex that is not an input or output vertex $d_{in}(v) = d_{out}(v)$.

\begin{definition}\label{def:legaldag}
\em An $input ~edge$ is an edge outgoing an input vertex, and an $output~ edge$  is  an edge  incoming an output vertex. 
\end{definition}
We will restrict our discussion to dags that fulfill the following conditions:
\begin{align*}
\text{Condition 1:} \quad & d_{out}(v) = 1 && \forall v \in In(G), \\
\text{Condition 2:} \quad & d_{in}(v)  = 1 && \forall v \in Out(G), \\
\text{Condition 3:} \quad & d_{in}(v) = d_{out}(v) && \forall v \in V \setminus \{ In(G) \cup Out(G) \}.
\end{align*}
\noindent 
Taking into account that every unitary operator that acts on $n$ qubits can be decomposed into a sequence of many single- and two-qubit quantum gates \cite{1995elementary}, for such a specific case, Condition 3 will be replaced by:
\begin{align*}
\text{Condition 3'}: d_{in}(v) = d_{out}(v) ~= ~ 2 &&  \forall v \in V \setminus \{ In(G) \cup Out(G)\}.    
\end{align*}

\begin{definition}\label{def:legal dag}
\em A  dag $G(V,E)$, $E \neq \emptyset$, satisfying Conditions 1, 2 and 3 is termed $legal$. A  dag satisfying Conditions 1, 2 and 3' is termed 2-\textit{legal} (clearly a dag that is 2-\textit{legal} is also \textit{legal}).   
\end{definition}  
Hereinafter, all the analyzed graphs are \textit{legal} dags.
\begin{theorem}\label{theorem:dag3}
Let $G$ be a \textit{legal} dag. Then there exists an integer $t>0$ such that\\ (a)  $|In(G)|=|Out(G)|=t$.\\ 
(b) There are $t$ edge-disjoint paths connecting an input vertex and an output vertex. \begin{ignore}
(Note: such a $t$ exists according to Theorem  $\ref{theorem:dag3} (a)$). Then we can order the vertices in $In(G)$ and $Out(G)$ as 
$In(G)=\{x_1, x_2, ..., x_t\}$, $Out(G)=\{y_1, y_2, ..., y_t\}$ such that there is a directed path from $x_i$ to $y_i$ for every $1 \leq i \leq t$.
\end{ignore}
\\
Proof: For (a) see Appendix \ref{appedix:lemma1}, for (b) see Appendix \ref{proof:Thorem 1}.
\end{theorem}

Theorem \ref{theorem:dag3}(a) emphasizes that a \textit{legal} dag represents unitary quantum channel, thus it is a valid quantum program, where we measure all incident qubits and there is no information leakage. Theorem \ref{theorem:dag3}(b) builds upon the fact that  each qubit is being measured and consequentially one can find a direct path between a qubit and a measurement device. As in the quantum knitting protocol, cutting a qubit  yields a mid-circuit measurement in the first partitioned graph and a corresponding pre-selected state in the second, so all the qubits are being used. 

Note that we have been using the aforementioned \textit{legal} dag definition, which is a valid representation of a quantum circuit. The admissible operations on \textit{legal} dags will be defined in the next section. 

\section{Operations on legal dags}\label{sec:operations on dag} 

Given a dag $G(V,E)$, a regular deletion of an edge $e$ results in the dag $G=(V,\tilde{E})$, where $ \tilde{E}=E \backslash \{e\}$.  In this work, we use a more elaborate mechanism, which we term $duplication$.

\begin{definition}\label{def:duplication}
 \em Let $G(V,E)$ be a dag,   and $E' \subseteq E$. The graph  $G_{E'}$ is defined as follows:
If $|E’| = 0 $ then $G_{E’}=G$.
If $|E’|=1$, that is $E’=\{e=(a,b)\}\in E$,  Then  $G_{E’} = (V \cup \{x,y\}, E \setminus \{e\} \cup \{(a,x),(y,b) \}$, where $ x,y \notin V$ and $x \neq y$. In this case, we denote $G_{E’} $ as $G_e$. 
This operation is termed \em duplication \em of  $e$, and $G_e$ is the {\em duplicated} graph. 
  If $|E’|>1$ then the graph $G_{E’}$ is obtained by duplicating  the edges in $E'$. Clearly, all orders  in which we duplicate these edges in $E'$  result in an isomorphic graph  $G_{E’}$.
\end{definition}

\begin{figure}[h!]
    \centering
    \begin{subfigure}[b]{1\textwidth}
        \centering
        \includegraphics[width=\linewidth]{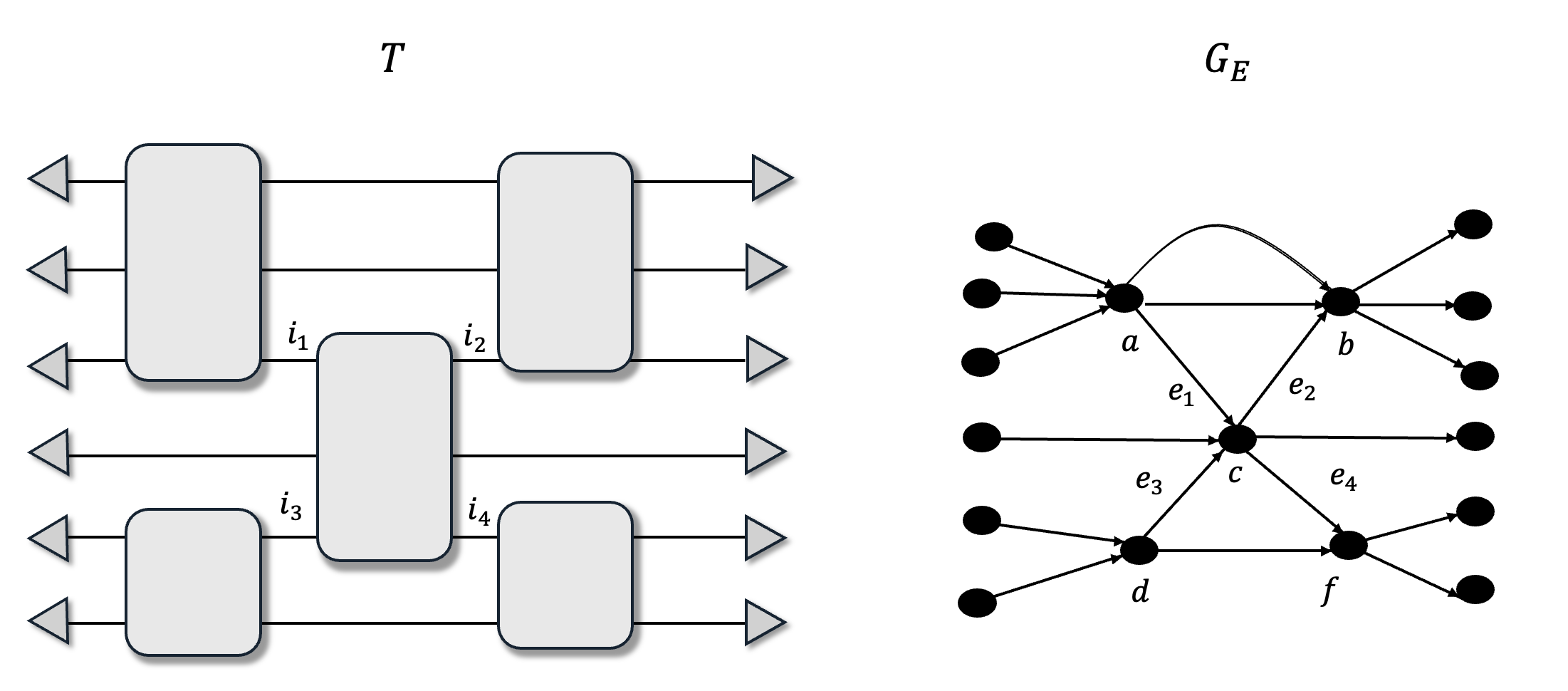}
        \caption{}
        \label{def7_Fig1}
    \end{subfigure}
    \hfill
    \begin{subfigure}[b]{1\textwidth}
        \centering
        \includegraphics[width=\linewidth]{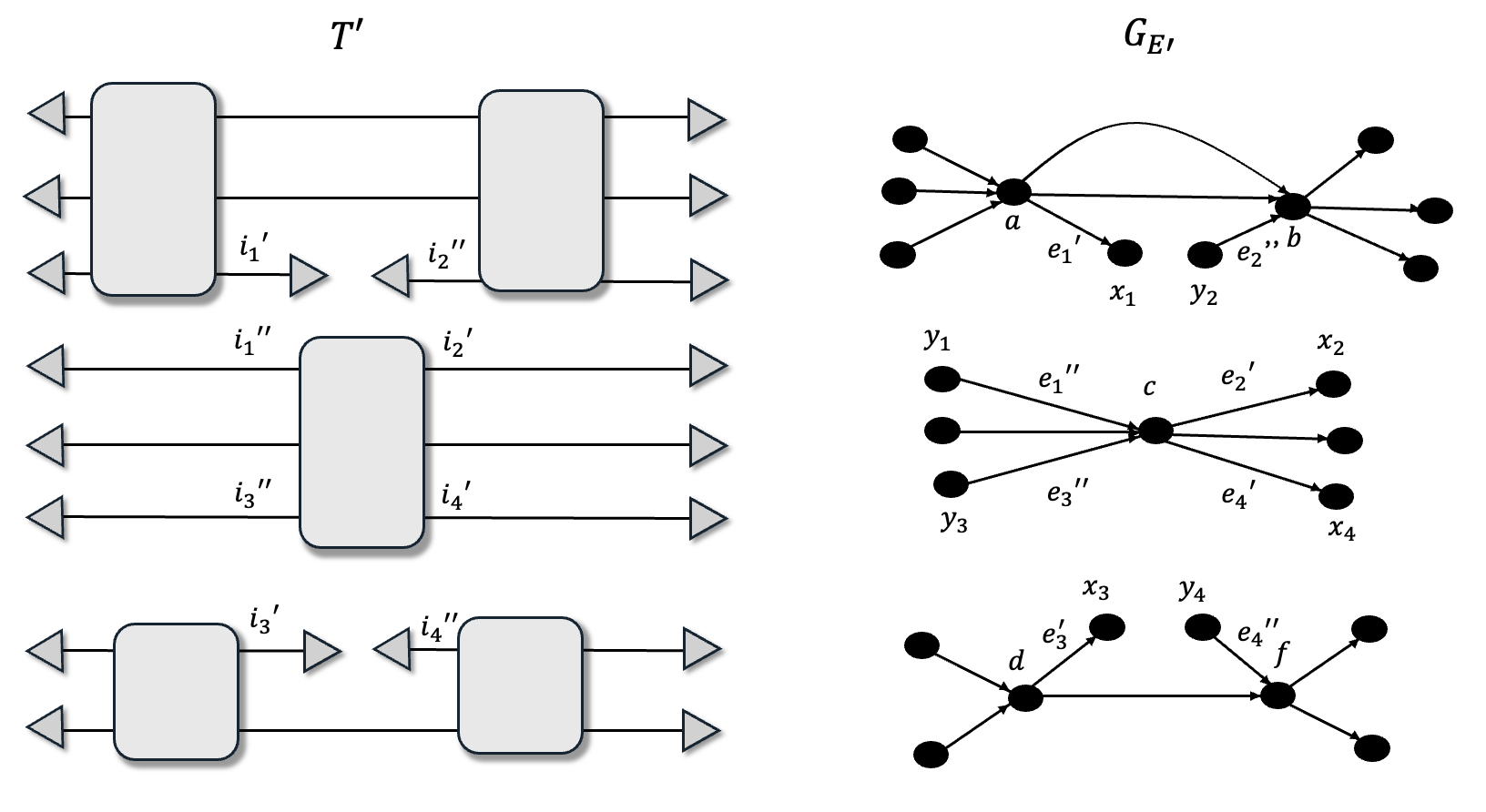}
        \caption{}
        \label{def7_Fig2}
    \end{subfigure}
    \caption{An example demonstrating $G_{E'}$ of Definition \ref{def:duplication}, for $|E'| =4$. (a) Simple tensor network $T$, and the connected graph ($G_E$), that represent the tensor network. The desired edges that we wish to delete are $E ' =\{e_1, e_2, e_3,e_4\} \subseteq E$. (b) Tensor network $T'$ and $G_{E'}$, obtained from the graph duplication, the set $E' = \{e_1, e_2,e_3,e_4\}$, the additional vertices $V' = \{x_1,y_1, x_2, y_2, x_3, y_3, x_4, y_4\} \notin V$ and the additional edges.}
    \label{Fig:definition_7}
\end{figure}
Note: In our graph formulation, each timelike (wire) cut is represented by an \emph{edge duplication} operation (Definition~\ref{def:duplication}). Hence we use “number of duplications” and “number of wire cuts” interchangeably when referring to the cut budget $\beta$.

The following lemma is easily derived from the definitions, and its proof is omitted:  
\begin{lemma}\label{lemma:dag1}
\em Let $G(V,E)$ be a dag,  $e \in E$, and $E' \subseteq E$. Then $G_e$ and $G_{E'}$ are both dags. In particular,
$In(G_e)=In(G) \cup \{y\}$, and 
$Out(G_e)=Out(G) \cup \{x\}$. 
\end{lemma}
An example for the case of $E’=4$: $T$ and $T'$ are tensor networks before and after the duplication protocol and the graphs associated with them are shown in  Fig.~\ref{Fig:definition_7}, where four edges are duplicated in the original graph in Fig.~\ref{def7_Fig1}; $E' = \{e_1 = (a,c),e_2 = (b,c),e_3 = (c,d),e_4 = (c, f)\}$, and the duplicated  graph  $G_{E'}$ is depicted in Fig.~\ref{def7_Fig2}.

\begin{definition}\label{def:tee and forest}
\em
Given a dag $G=(V,E)$, $G'=(V,E')$ is the undirected graph where $E'=E$; namely, each directed edge $(a,b)\in E$ is viewed as an undirected edge in $E'$ \footnote{Since we use only dags, $G'$ cannot contain parallel edges} (informally, the undirected graph $G'=(V,E')$ is constructed from $G$ by ignoring the directions of the edges).  $G'$  is termed the {\it underlying undirected graph} of $G$. A $connected ~dag$ is a dag whose underlying undirected graph is connected. A $tree~ dag$ is a dag whose  underlying undirected graph is a tree, and a $forest~ dag$ is a dag whose  underlying undirected graph is a forest (recall that a forest is a collection of trees).
    
\end{definition}

\noindent 
 For example, the dag where  $E = \{(a,b),(a,c)\}$ is connected, but the dag where $E = \{(a,b),(c,d)\}$ is not connected.

\begin{definition}\label{def:lod10}
 \em Given a dag $G$, its underlying undirected graph $undirected(G)$ is partitioned into connected components. Each component is a connected undirected graph $undirected(G')$, for some  subgraph $G'$ (see example in Appendix.~\ref{ref:example9})  of $G$. $G'$ is termed a {\em component} of $G$. A \textit{cluster} of $G$  is a subgraph of $G$ that consists of one or more components.    
 
\end{definition}
 
\begin{lemma}\label{lemma:number of components}
\em Let $G(V,E)$ be a dag and $e \in E$. Then  the number of components in $G_e$ is either unchanged or increased by one.  
\end{lemma}
This follows from the definitions, and its proof is omitted.
Recall that, according to Theorem \ref{theorem:dag3}, in each component of a legal dag $G$  there is a path from a vertex in $In(G)$ to a vertex in $Out(G)$.
However, by duplicating an edge, this property can be violated (for example, if the dag $G$ contains only one edge $e = (a,b)$, its duplicated graph $G_e$ is not acceptable.)
We will use only $acceptable$ duplications, defined as follows: 

\begin{definition}
\em
Let $G(V,E)$ be a dag, and $E' \subseteq E$. Then the duplicated graph $G_{E'}$ is $acceptable$ if  each of its components contains  a path connecting a vertex $a \in In(G)$ and a vertex $b \in Out(G)$.  
\end{definition}

\noindent

\begin{lemma}\label{lem:acceptable}
Let $G=(V,E)$ be a legal dag, and $e \in E$. Then
\begin{itemize}
    \item if $e$ is an input edge then $G_e$ is not acceptable.  
    \item if $e$ is an output edge then $G_e$ is not acceptable. 
    
\end{itemize}
\end{lemma}
\begin{proof}
It is clear that in either case, one of the resulting subgraphs does not contain an original input vertex or an original output vertex, thus the resulting dag is not acceptable.     
\end{proof}

\section{Graph Duplication}\label{sec:graph duplication}

Given a legal dag $G(V,E)$, and integers $k,\alpha \geq 1$, $\beta \geq 0$, we want to duplicate at most $\beta$ edges,
such that the resulting graphs will be partitioned into at most $\alpha$ clusters, each with at most $k$ inputs and $k$ outputs (recall that a cluster has one or more components, see Definition \ref{def:lod10}). Moreover, we require that each of these clusters will be acceptable.  
Formally, given a legal dag $G(V,E)$, we want to find a subset $E' \subseteq E$, $|E'| \leq \beta$, such that the resulting graph $G_{E'}$ will contain at most $\alpha$ disjoint acceptable subgraphs,  each including at most $k$ inputs and $k$ outputs.
Note that input graph $G$ is not necessarily connected, meaning that some subset of qubits does not interact with another subset of qubits, so one can simulate them separately.
\noindent
Formalized as a decision problem, we ask whether there is a solution to the following problem:

\vspace{5 mm}
\noindent
\begin{definition}\label{def:Graphup}
The Graph Duplication problem   $GD(G,k,\alpha,\beta)$.

\noindent \textit{Input:} A legal dag $G=(V,E)$ ,  integers $k,\alpha \geq 1$ and $\beta \geq 0$.

\noindent \textit{Question:}  Does there exist  $E' \subseteq E$, $|E'| \leq \beta$, such that all the components of  
$G_{E'}$ are acceptable, and $G_{E'}$ can be partitioned into clusters $G_1, G_2, \cdots, G_p$, $~p \leq \alpha$, satisfying
 $|In(G_i)~|= |Out(G_i)| \leq k$ for every $1 \leq i \leq p$?

\end{definition}

\begin{ignore}
    Note that, for Lemma \ref{lem:acceptable}, a solution $E'$ to an instance of problem $GD(G,k,\alpha,\beta)$ cannot include an input edge or an output edge. 
\end{ignore}
\section{The complexity of the GD problem}\label{Sec:resutls}
\subsection{Overview}
This section classifies the computational complexity of the GD problem under different restrictions on the input graph $G$ and on the cut budget $\beta$. First, we show that GD can be decided in polynomial time when $G$ is a connected legal dag and $\beta$ is constant. This holds even in the case where the input graph $G$ has a constant number of connected components (Theorem \ref{GD_COMPLEXITY}). Second, GD becomes NP-complete in all other cases; namely, when the number of components is not bounded or the number of duplications (cuts) $\beta$ is not bounded. The NP-completeness proof proceeds through a sequence of reductions: it is shown first for forests and $\beta = 0$ (Theorem \ref{th:beta=0}) and then for any constant $\beta$ (Theorem \ref{th:4beta>0}). The same two cases are then shown to be NP-complete also in 2-legal forests (Theorems \ref{th5:beta=0} and \ref{th:2legal, beta>0}). Last we show the problem is NP-complete for 2-legal trees and $\beta >0$ (Theorem \ref{th:connected}). In our analysis, we use a reduction from the 3-partition NP-complete problem \cite{garey1975complexity}, which is explicitly defined in Appendix \ref{Appendix:3partition}. The main focus of the problem will be on any general legal dag, and in particular on 2-legal dags.

\subsection{Connected $G$, constant $\beta$}

\begin{theorem}\label{GD_COMPLEXITY}
Problem $GD(G,k,\alpha,\beta)$ can be solved in polynomial time for a connected dag $G$, for any constant $\beta$. This holds even in the case where the input graph $G$ is composed of a constant number of connected components.   
\end{theorem}
\begin{proof}
Given a dag $G=(V,E)$, we first list all possible subsets of at most $\beta$ edges; this can be done in time  $O(|E|^1+|E|^2+\cdots |E|^{\beta})=O(|E|^{\beta})$. 
Then, for each of these subsets $E'$, we consider the dag $G_{E'}$. 
We first determine its 
components; this can be done in a time $O(|E'|)$. 
According to Lemma \ref{lemma:number of components}, 
the number of components  is at most $\beta+1$. We then examine all partitions of these components into at most $\alpha$ clusters.  But the number of clusters is clearly bounded by $\beta+1$.  
So we have to examine all possible partitions of these  components into at most $\beta +1$ clusters. The number of these cases  is  $O((
\beta+1)^{\beta +1})$.
Each such cluster $C$ is then checked in order to determine whether it is acceptable. This is done in time $O(|C|)$. 
\noindent
We have thus established a polynomial time solution to problem $GD(G,k,\alpha,\beta)$. 
The extension for the case where $G$ is composed of a constant number of components is straightforward.
\end{proof}

\subsection{Forest $G$, $\beta = 0$}

\begin{theorem}\label{th:beta=0}
Problem Graph duplication $GD(G,k,\alpha,\beta)$ is NP-complete for $\beta = 0$, even when $G$ is a forest dag. 
\end{theorem}

\begin{proof}

We show a polynomial time reduction from the 
3-partition problem to 
$GD(G,k,\alpha,0)$.

\begin{figure}[ht]
    \centering
    \includegraphics[width=0.9\linewidth]{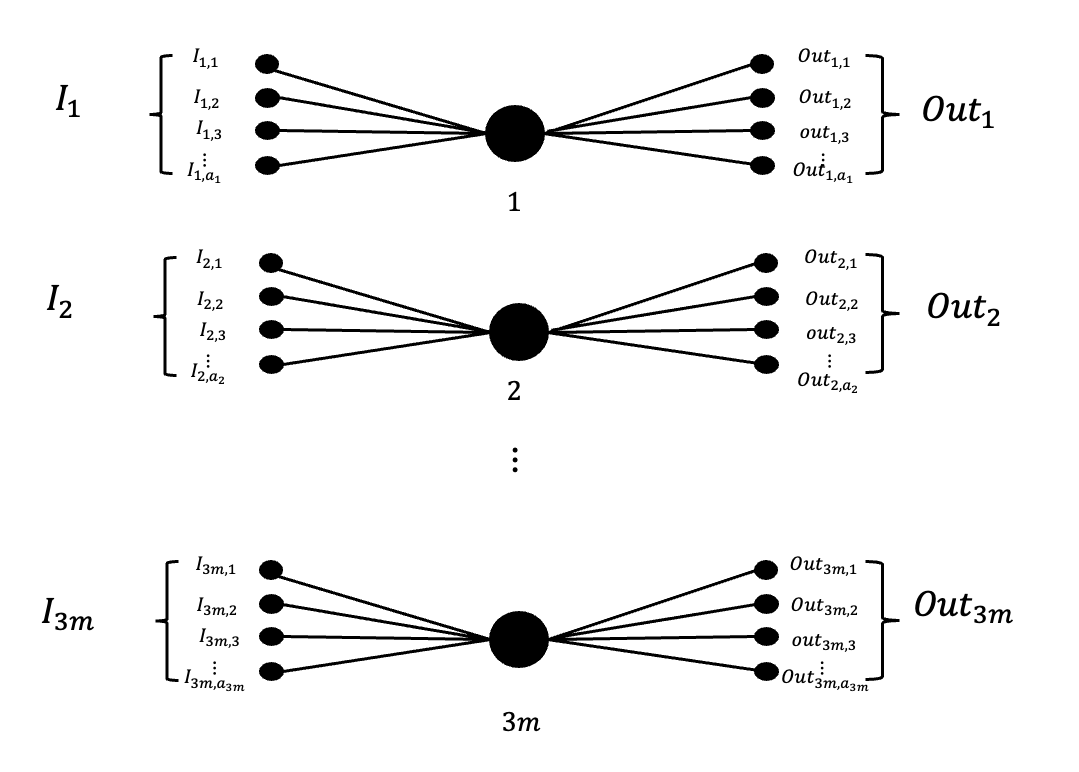}
    \caption{The case $\beta = 0$. The dag shown is $G^0$.}
    \label{Fig:beta=0}
\end{figure}

\noindent
Given an instance $I$ of 3-partition - that is, a set of $3m$ positive integers $a_1,a_2,...,a_{3m}$, $\sum_{i=1}^{3m} a_i= mB$, such that  $\frac{B}{4} < a_i < \frac{B}{2}$ for every $1 \leq i\leq 3m$ - 
we define the following instance of   $GD(G,k,\alpha,0)$.
The graph is $G=G^0=(V,E)$. The set of vertices $V$ is
$V=\cup_{i=1}^{3m}In_i \cup\{1,2,\cdots,3m\}
\cup_{i=1}^{3m}Out_i$, where $In_i=\cup_{j=1}^{a_i}\{in_{i,j}\}$ and $Out_i=\cup_{j=1}^{a_i}\{out_{i,j}\}$, for $1 \leq i \leq 3m$.
The set of edges $E$ is 
$E=\cup_{i=1}^{3m}\cup_{j=1}^{a_i}\{(in_{i,j},i),(i,out_{i,j})\}$. (See Fig~
\ref{Fig:beta=0}.)

We set $k=B$ and $\alpha = m$. Since $\beta = 0$, no edge was duplicated, so the resulting graph is clearly acceptable. In addition, the reduction can be done in time $O(m)$, and is thus polynomial in the input size.

\noindent
We have to show that there is a solution to the given instance $I$ of  3-partition iff there is a solution to the  problem $GD(G^0,B,m,0)$. 

Assume there is a solution to the 3-partition problem with input $I$. 
We show that in this case there is a solution to $GD(G^0,B,m,0)$; that is, a solution where we do not need to duplicate any edge. For each of the $m$ triples $(\delta_1, \delta_2, \delta_3)$ in the solution - that is $\delta_1, \delta_2, \delta_3 \in I$ and  $\delta_1+ \delta_2 + \delta_3 = B$ - the subgraph of $G^0$ composed of $V=In_{\delta_1}\cup In_{ \delta_2}\cup In_{\delta_3}\cup\{\delta_1, \delta_2, \delta_3\}\cup Out_{\delta_1}\cup Out_{ \delta_2}\cup Out_{\delta_3}$, together with all edges, contains exactly $B$ inputs and $B$ outputs. 
Clearly, there are exactly $B$ paths connecting an original input vertex to an original output vertex. This partitions $G^0$ to exactly $m$ parts as required.

Conversely, if there is a solution to $GD(G^0,B,m,0)$ then $G^0$ must be partitioned into subgraphs, containing the basic subgraphs with  $\delta_1, \delta_2, \delta_3$ inputs, such that $\delta_1+ \delta_2 + \delta_3 = B$. This establishes a solution to the given instance $I$ of the 3-partition problem.  
\end{proof}

\subsection{Forest $G$, $\beta>0$ constant}\label{Subsec:Gforestbetabigzero}

\begin{theorem}\label{th:4beta>0}
Problem Graph duplication   $GD(G,k,\alpha,\beta)$ is NP-complete for any  constant $\beta>0$, even when $G$ is a forest dag. 
\end{theorem}
The proof is an extension of the proof of Theorem \ref{th:beta=0} (see Appendix \ref{theorem4sketch}).


\subsection{Forest $G$, 2-legal dags, $\beta =0$}
Our motivation for the specific proof of the 2-legal dag defined in Condition $3'$ is that any multi-qubit gate can be decomposed into a sequence of one- and two-qubit gates, involving at most two-qubit interactions \cite{1995elementary}.

We now extend this NP-completeness result to 2-legal dags, that is, dags that satisfy Condition 3'.

\begin{theorem}\label{th5:beta=0}
Problem Graph duplication   $GD(G,k,\alpha,\beta)$ is NP-complete for 2-legal dags $G$, for $\beta=0$, even when $G$ is a forest dag.  
\end{theorem}
The proof appears in Appendix \ref{appendix:theorem5sketch}.

\subsection{Forest $G$, 2-legal dags $\beta >0$ constant}

\begin{theorem}\label{th:2legal, beta>0}
Problem Graph duplication   $GD(G,k,\alpha,\beta)$ is NP-complete for 2-legal dags $G$, for any constant  $\beta>0$, and the underlying undirected graph of $G$ is a forest. 
\end{theorem}
The   proof is in Appendix \ref{appendix:Theorem 6 sketch }.

\subsection{2-legal connected dags, non constant $\beta$ }

\noindent
\begin{theorem}\label{th:connected}
Problem Graph duplication $GD(G,k,\alpha,\beta)$ is NP-complete for connected legal dags $G$, even in the case where $G$ is a 2-legal tree dag. 
\end{theorem}

\noindent{Note: In this case $\beta $ is not  a constant (in the case where $\beta$ is constant the problem can be solved in polynomial time according to Theorem \ref{GD_COMPLEXITY}}).
\begin{proof}

We extend the proof of Theorem  \ref{th:beta=0}.

Given an instance $I$ of 3-partition, we construct  the graph $\tilde G$ depicted in Fig.~
\ref{Fig:connected}.

We show that there is a solution to the 3-partition problem  $I$ iff there is a solution to problem $GD(\tilde{G},B+3,4m-1,6m-2)$ (recall that in this problem  we need  to duplicate at most $6m-2$ edges, and partition $\tilde G$ to at most $4m-1$ parts, each with at most $B+3$ inputs). A crucial point in this reduction is that, though $\tilde G$ is connected, we do not require that each of these parts will be connected. Assume there is a solution to $I$. Then we can duplicate all the edges $\{e_1,\,e_1^{\prime},\,\ldots,\,e_{3m-1},\,e_{3m-1}^{\prime}\}$.
$\noindent $ All $3m-1$ components containing $A_i$  and $B_i$ 
(see Fig~\ref{Fig:connected}) contain exactly $B+3$ inputs each. The other components can be grouped into $m$ triples $A_i,A_j,A_k$, according to the solution of the 3-partition problem $I$. Each such component will have  $(a_i +1)+(a_j+1)+(a_k+1) =a_i+a_j+a_k+3=B+3$ inputs, thus this part will have exactly $B+3$ inputs. We have thus established a solution to  
$GD(\tilde{G},B+3,4m-1,6m-2)$.

Conversely, assume there is a solution to $GD(\tilde{G},B+3,4m-1,6m-2)$.
It is clear that each of the $A_i$'s must be in a separate component. Thus it is clear that all edges 
$\{\,e_1,\,e_1^{\prime},\,\ldots,\,e_{3m-1},\,e_{3m-1}^{\prime}\,\}$ must be duplicated. As there are exactly $6m-2$ such edges, it follows that no other edge can be duplicated.
Each component containing $A_i$ has exactly $B+3$ inputs. 

We thus have a total of $3m-1$ parts. By Lemma \ref{lem:acceptable} it follows that we have at most $4m-1$ parts in the partition. Therefore, the $3m$ parts containing the $In_i$'s must be partitioned into at most $m$ parts, each with  at most $B+3$ inputs. 
Note that all elements in each $In_i$ are in the same partition, according to Lemma \ref{lem:acceptable}.
One component with the $In_i$ as inputs contains $a_i +1$ inputs. 

Since $\frac{B}{4} < a_i < \frac{B}{2}$, we have 
$\frac{B}{4} +1 < a_i +1 $.
Thus, the sum of four or more of the $a_i$'s is more than B+4. So each of the partitions contains at most three such sets $In_i$. But, since there are $3m$ such sets,  if any  partition has less than 3 then there will be a partition with more than 3, a contradiction. Thus, each partition has exactly 3 such sets $In_i$. If they correspond to the elements $a_i,a_j,a_k$, then $(a_i+1)+(a_j+1)+(a_k+1) \leq B+3$, 
thus $a_i+a_j+a_k \leq B$.

Summing for all $m$ triples, we get that $\sum_i^{3m} a_i \leq mB$. But, since $\sum_i^{3m} a_i = mB$,
it follows that for each of these components $a_i+a_j+a_k = B$. We have thus established a solution to $I$.

Extending the result to the case of 2-legal dags is done by using the technique presented in the proof of Theorems \ref{th5:beta=0} and \ref{th:2legal, beta>0}. 

\begin{figure}[ht]
    \centering
    \includegraphics[width=0.9\linewidth]{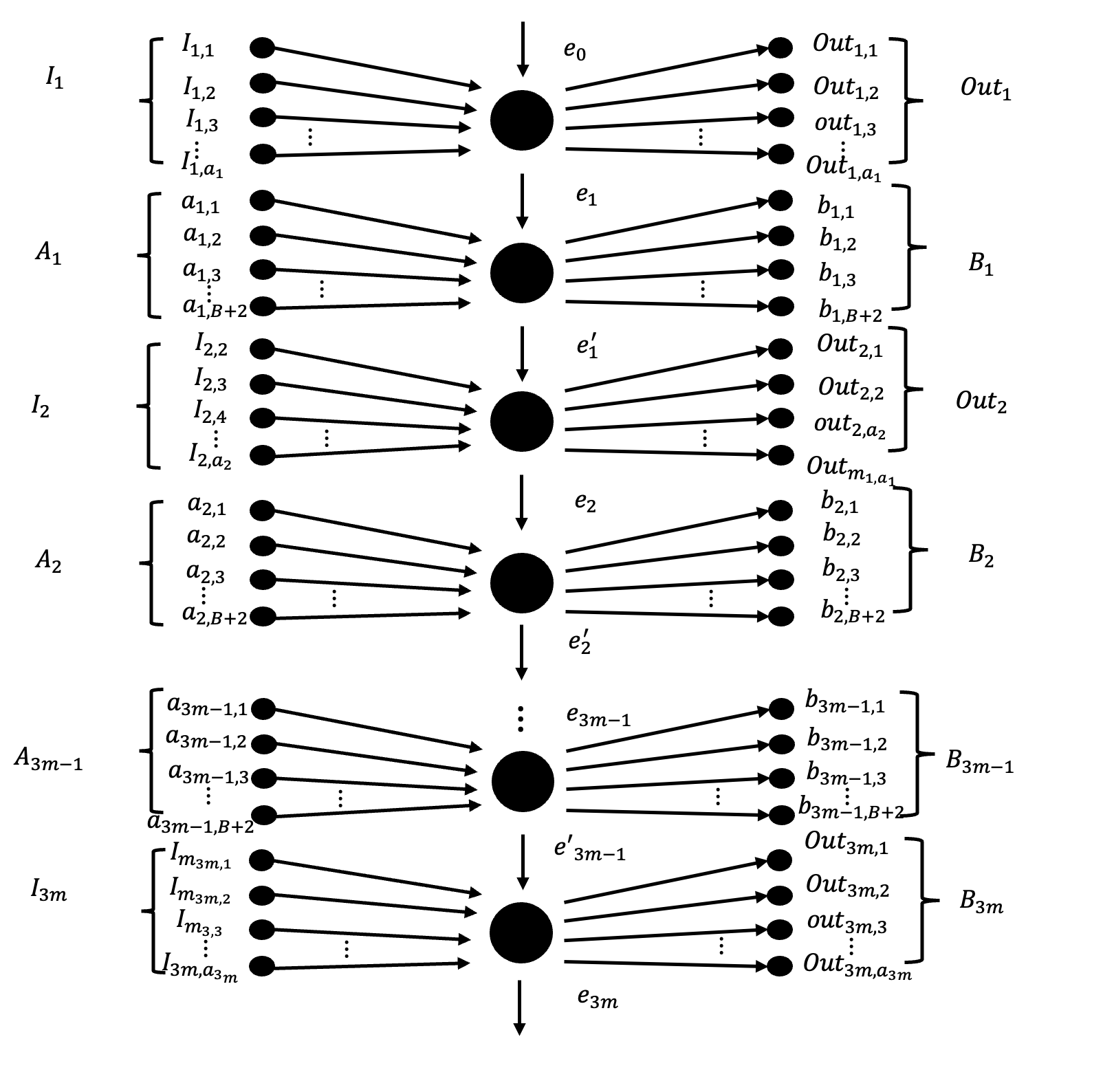}
        \caption{The connected dag $\tilde{G}$ used in the proof of Theorem \ref{th:connected}. }
    \label{Fig:connected}
\end{figure}
\end{proof}

\section{SMT solver}\label{Sec:smt_solver}
Motivated by the analysis in \cite{brandhofer2023optimal}, we propose in this section an SMT-based solution for the circuit cutting problem taking into account our aforementioned definitions.
The SMT solver (Microsoft Z3) was constructed based on our GD problem (Definition \ref{def:Graphup}), where the solver  attempts to minimize the number of duplications $\beta$. If the  SMT solver does not find a solution, the program will end without any partition. The SMT formulation provides an exact optimization framework for the encoded GD problem, suitable for modest instance sizes (and relatively small values of $\beta$), although worst-case runtime remains exponential.
The exact constraints are:
\begin{itemize}

\item The set $O = \{\, o_{v,p} \mid v \in \mathcal{V},\ p \in \mathcal{P} \,\}$, with each $o_{v,p} \in \{0,1\}$ indicating whether vertex $v$ is in partition $p$.

\item The set $C = \{\, c_e \mid e \in E \,\}$, with each $c_e \in \{0,1\}$ indicating whether edge $e$ is cut.

\item For any $v$ and $p \neq p'$, $o_{v,p} + o_{v,p'} \le 1$. In other words, for each $v$, $\sum_{p \in \mathcal{P}} o_{v,p} = 1$.

\item For each partition p$, \exists v_i \in \mathrm{In}(G),\ \exists u_j \in \mathrm{Out}(G)$: there is a path $\mathrm{p}(v_i,u_j)$, where $\mathrm{p}(v_i,u_j)$ implies that there is a path from $v_i$ to $u_j$ using only vertices and edges in partition $p$.

\end{itemize}

\begin{figure}
     \centering
     \begin{subfigure}[b]{0.25\textwidth}
         \centering
         \includegraphics[width=\textwidth]{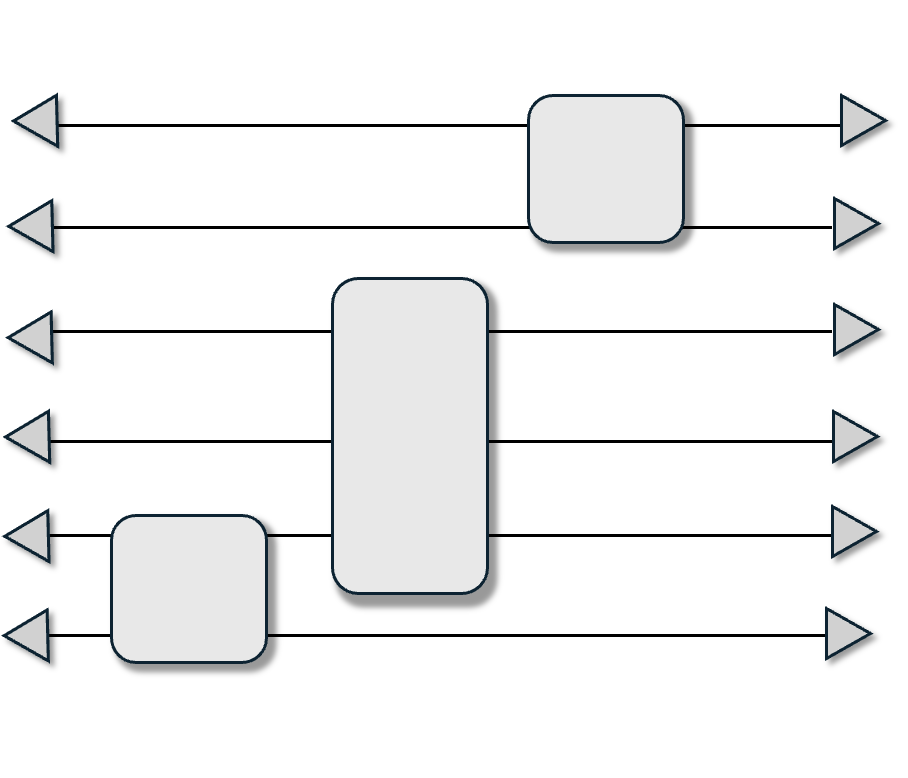}
         \caption{}
         \label{fig:already_sep}
     \end{subfigure}
     \hfill
     \begin{subfigure}[b]{0.3\textwidth}
         \centering
         \includegraphics[width=\textwidth]{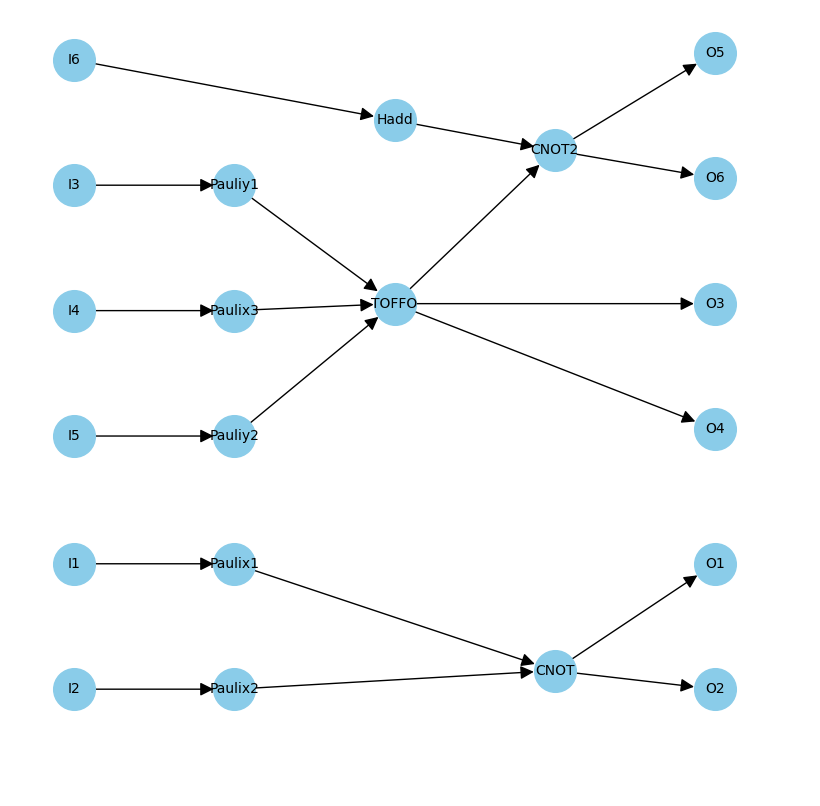}
         \caption{}
         \label{Fig:firstgraph_sep}
     \end{subfigure}
     \hfill
     \begin{subfigure}[b]{0.3\textwidth}
         \centering
         \includegraphics[width=\textwidth]{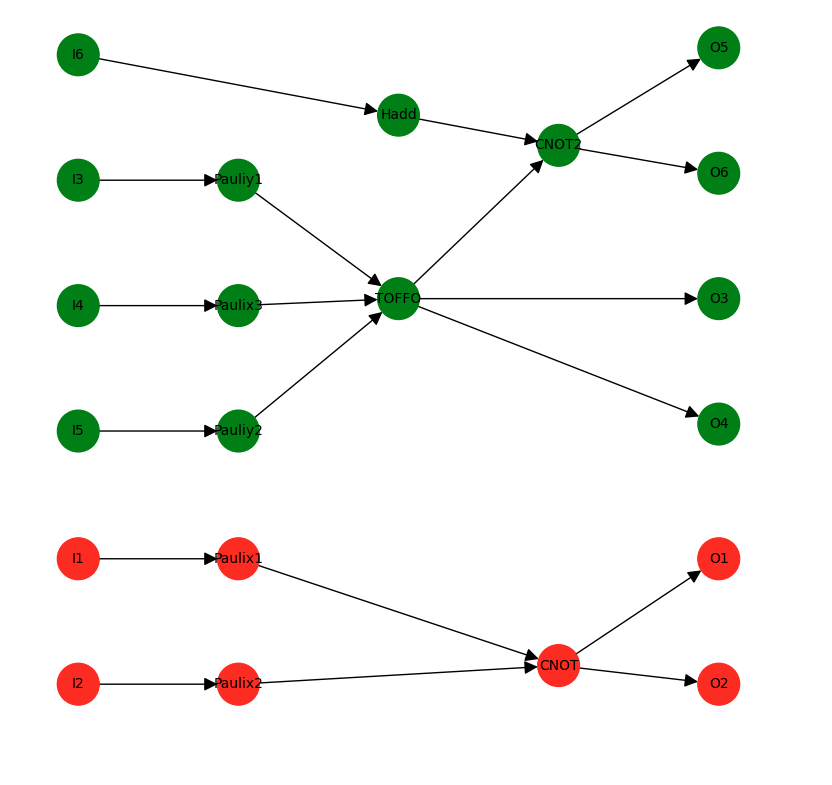}
         \caption{}
         \label{Fig:Result_sep}
     \end{subfigure}
        \caption{(a) Arbitrary quantum algorithm represented in the circuit model, each block is unitary operator with the rank of the number of inputs and outputs qubits. (b) The graph representation of the circuit in (a), which is now a legal dag. (c) After applying the constraints, the SMT solver returns two disjoint circuits, highlighted in different colors.}
        \label{Fig:three graphs}
\end{figure}

The proposed SMT solver (see \cite{smt_circuit}) takes as an input  a legal  dag, a maximum per-partition qubit budget 
$Q$, and an upper bound on the number of partitions $P_{max}$
max. The legal dag G contains primary inputs $In(G)$ (nodes with no predecessors) and  primary outputs $Out(G)$ (nodes with no successors).  Having a legal dag implies that  $|In(G)|=|Out(G)|$. The solver then tries $P=1,2,…,P_{max}$  and stops at the first feasible partitioning.
The variables, based on the constraints that were introduced earlier, are: a partition index for each vertex, the number of primary inputs per partition, the number of crossing edges entering each partition, and the total number of duplications, which we minimize. The SMT includes an optional requirement demanding each partition to be fully connected, and for that we use a lightweight breadth-first search (BFS) scaffold to enforce connectivity within each partition (within the ``3 partition'' reductions this requirement is not necessary).
In the program, each vertex is assigned a partition. For each partition the total ``qubits'' accounted for are the primary inputs  plus the incident crossing edges that enter the partition, empty partitions are not allowed. Optionally, one can require that each partition contains at least one $In(G)$ and $Out(G)$ of the same qubit.
When requiring each partition to be internally connected, the BFS-like scaffold is defined as follows: in each partition, exactly one node is selected as a root, and every other node in the partition is assigned a strictly positive distance from it. For any edge whose endpoints lie in the same partition, the assigned distances differ by at most 1, and every non-root node has an in-partition neighbor at distance one or less. These constraints guarantee that the subgraph induced by each partition is connected.

\begin{figure}
     \centering
     \begin{subfigure}[b]{0.25\textwidth}
         \centering
         \includegraphics[width=\textwidth]{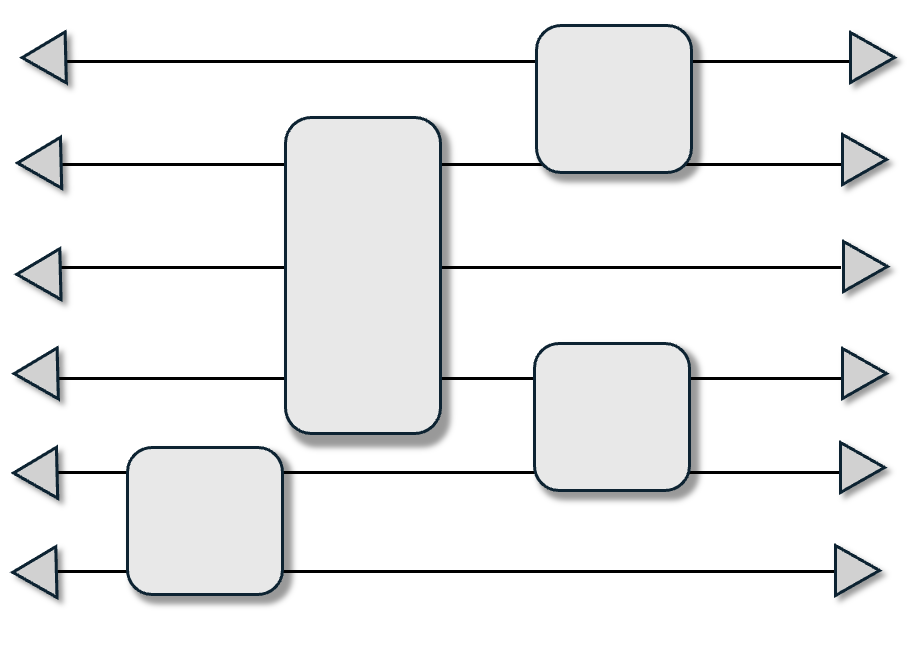}
         \caption{}
         \label{fig:second_smt_example}
     \end{subfigure}
     \hfill
     \begin{subfigure}[b]{0.3\textwidth}
         \centering
         \includegraphics[width=\textwidth]{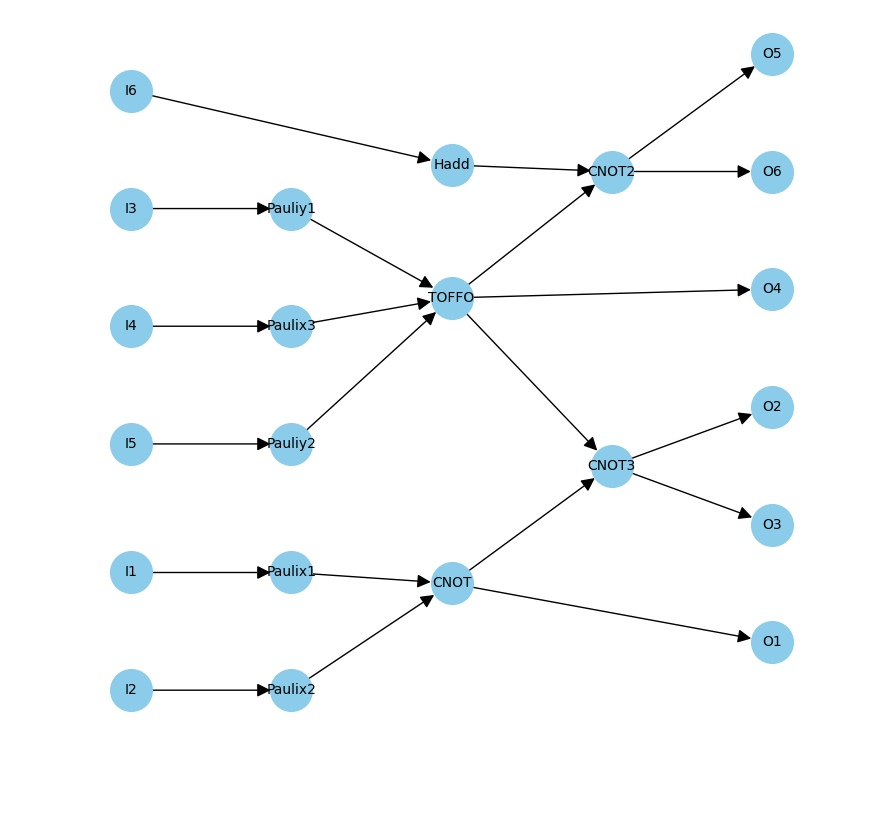}
         \caption{ }
         \label{fig:second_graph_}
     \end{subfigure}
     \hfill
     \begin{subfigure}[b]{0.3\textwidth}
         \centering
         \includegraphics[width=\textwidth]{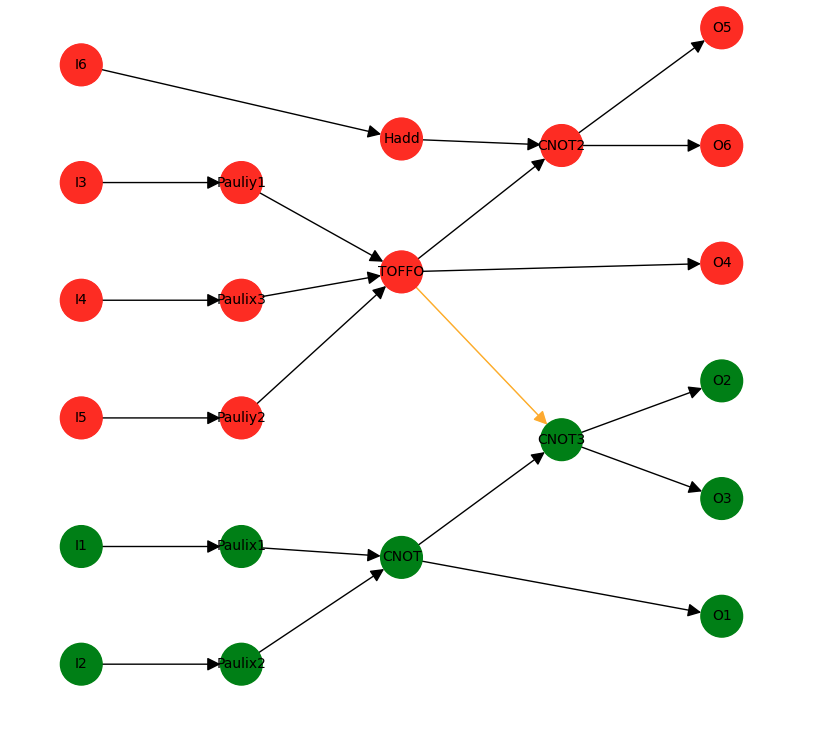}
         \caption{}
         \label{fig:second_graph_sol}
     \end{subfigure}
        \caption{(a) Quantum algorithm with induced entanglement, decomposition necessitates qubit cuts. (b) The graph representation of the circuit. (c) SMT solution for the cutting protocol, yielding two separable subgraphs (red and green); the yellow edge indicates the qubit cut. }
        \label{fig:three graphs}
\end{figure}

The first example depicted in Fig.~\ref{fig:already_sep}, is a unitary operation that can be written as a tensor product of local unitaries. The SMT solver visualization is presented in Fig.~\ref{Fig:firstgraph_sep}, visualized using our Python implementation (NetworkX). The details of the simulation are as follows. We asked the SMT to give at most two clusters with at most 4 qubits in each. The SMT solver solution depicted in Fig.~\ref{Fig:Result_sep}, where each cluster has a different color. As was implied above, the SMT solver found partitions without duplicating any edge. Now consider the next test case, in which a feasible solution exists only if duplication is allowed. Repeating the same procedure as in the previous example, the entangled circuit shown in Fig.~\ref{fig:second_smt_example} is mapped to the graph in Fig.~\ref{fig:second_graph_}, and the resulting solution is shown in Fig.~\ref{fig:second_graph_sol}. The duplicated edge is highlighted in yellow.

\section{Discussion}\label{Sec: DISCUSSION}
We introduced a graph-theoretic abstraction of timelike quantum circuit cutting and used it to define the graph duplication problem GD.  
We showed that the GD problem can be solved in polynomial time when $\beta$ is constant and $G$ has a  constant number of connected components.
Then we showed that the GD problem becomes NP-complete in all other cases; namely, when the number of components is not bounded or the number of duplications (cuts) $\beta$ is not bounded.
We also presented an SMT formulation aligned with the GD constraints. The SMT solver is intended as a practical optimization tool for moderate-size instances, not as a polynomial-time algorithm for the general problem: like other exact combinatorial solvers, it can still exhibit exponential worst-case runtime. Interesting questions for future research could be to characterize intermediate regimes in which $\beta$ grows with the instance size, for example $\beta = O(\log n)$, and to compare the present abstraction directly with hardware-aware circuit-cutting heuristics used in existing toolchains.
\section*{Acknowledgements} This work was supported by the European Union’s Horizon Europe research and innovation programme under grant agreement No. 101178170, by the Israel Science Foundation under grant agreement No. 2208/24 and by the Israeli Council for Higher Education through ``QERNEL'' and the ``Interdisciplinary Center for the Theory of Quantum Computing''.

\bibliographystyle{ieeetr}

\bibliography{references.bib}

\newpage

\appendix
\section{Proof extensions}

\subsection{Proof of Theorem \ref{theorem:dag3} (a)}\label{appedix:lemma1}

\begin{proof}

\noindent
In any directed graph $G=(V,E)~~~$ $\sum_{v \in V} d_{in}(v)=\sum_{v \in V} d_{out}(v) = |E|.$
Hence
$ \sum_{v \in {V \setminus (In(G) \cup Out(G))} }d_{in}(v) + \sum_{v \in In(G)}  d_{in}(v)
+ \sum_{v \in Out(G)} d_{in(v)}$
$=\sum_{v \in {V \setminus (In(G) \cup Out(G))} }d_{out}(v) + \sum_{v \in In(G)}  d_{out}(v)
+ \sum_{v \in Out(G)} d_{out(v)}$

Since the sums $ \sum_{v \in {V \setminus (In(G) \cup Out(G))}} $ on both sides of the equation are equal by Condition 3 above,  we have
  $\sum_{v \in In(G)}  d_{in}(v)
+ \sum_{v \in Out(G)} d_{in(v)} 
= \sum_{v \in In(G)}  d_{out}(v)
+ \sum_{v \in Out(G)} d_{out(v)}.$

\noindent Since $\sum_{v \in In(G)}  d_{in}(v)=\sum_{v \in Out(G)} d_{out(v)} = 0$ by the definition of inputs and outputs, we have
$$  \sum_{v \in Out(G)} d_{in(v)} 
= \sum_{v \in In(G)}  d_{out}(v)
.$$

\noindent
By Conditions 1 and 2 above, the term on the left side is equal exactly to $|Out(G)|$, since each vertex in $Out(G)$ contributes exactly $1$ to the sum, while the term on the right side is equal exactly to $|In(G)|$, since each vertex in $In(G)$ contributes exactly $1$ to the sum. Hence
$|In(G)|=|Out(G)|$.
\end{proof}

\subsection{Proof of Theorem 1 (b) }\label{proof:Thorem 1}
\begin{proof}
The proof is by induction on $t$. 

\noindent {\it Base:} $t=1$. In this case  $In(G)=\{x_1\}$,$~Out(G)=\{y_1\}$. We follow a path starting at $x_1$. Whenever the path reaches a new vertex $u$, there is always another edge outgoing $u$ (due to Condition 3). Since the graph is finite, and since we never enter a previously visited vertex (since $G$ is a dag), this procedure must terminate at $y_1$. Thus, we established a path from $x_1$ to $y_1$. 

\noindent {\it Induction step:} We show that if the claim holds for $t$ then it holds for $t+1$, for every $t \geq 1$. 
\noindent 
Let $G$ be a legal dag  in which  $|In(G)|=|Out(G)|=t+1$. Take any vertex $u \in In(G)$. Applying the same algorithm as above, we start at $u$, and continue along edges until we end at a vertex $u' \in Out(G)$. We now delete the edges of this path from $E$, and get a new graph $G'=(V,E')$, in which $|In(G')|=|Out(G')|=t$. 
It is clear that  after removing the path from the dag  the resulting graph is a dag. Moreover, it is a legal dag, since: (1) we deleted one input vertex and one output vertex, (2)  in all other vertices along the path the in-degree and out-degree were decreased by one,  thus maintaining Condition 3, (3) no new input or output vertex were created, and (4) the in and out degrees of all vertices which are not on this path remain unchanged. By the induction hypothesis, there are $t$ edge-disjoint paths connecting an input vertex and an output vertex.
\end{proof}
\subsection{Definition \ref{def:lod10} -  Example}\label{ref:example9}
\begin{figure}
\centering
    \hspace{-2.5cm} 
    \includegraphics[width=0.5\linewidth]{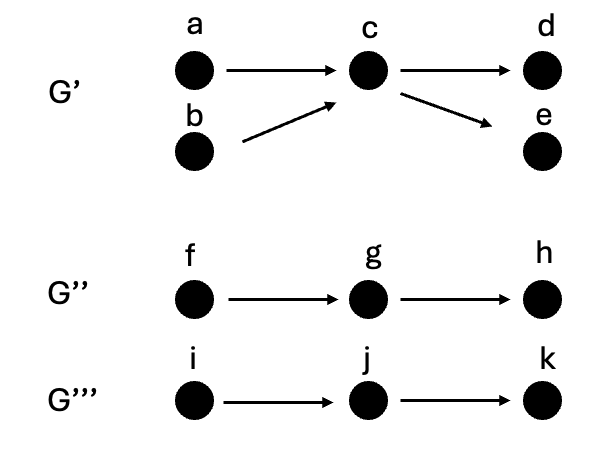}
    \caption{$G', G'', G''$ each contains one component while $G'' \cup G'''$ contains two.}
    \label{fig:componants examples}
\end{figure}

The dag G contains three
components $G', G'', G'''$. $G'$ is a cluster that contains one component, and $G'' \cup G'''$ is a cluster that contains two components. Fig.~\ref{fig:componants examples} illustrates these components.

\subsection{Proof  of Theorem 4 }\label{theorem4sketch}

\begin{figure}[ht]
    \centering
    \includegraphics[width=0.5\linewidth]{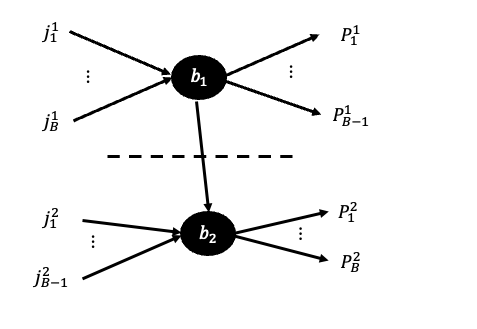}
    \caption{The case  $\beta > 0$. The dag shown is  $G'$, $\beta$ copies of which are added  to $G^0$.}
    \label{Fig:beta>0}
\end{figure}

The proof is an extension of that of Theorem \ref{th:beta=0}. 
Given an instance $I$ of 3-partition, we construct an instance  $GD(G^{\beta},B,m+2\beta,\beta)$;
$G^{\beta}$
includes the dag  $G^0$ (defined in the proof of Theorem \ref{th:beta=0}), to which we add  $\beta$ copies of the dag $G'=(V',E')$, where
 $V'=\{ a_{1,1},a_{1,2},\cdots,a_{1,B}, a_{2,1},a_{2,2},\cdots,$
 $a_{2,B-1}, t_1,t_2,b_{1,1},b_{1,2},\cdots,$
$b_{1,B-1}, b_{2,1},$ $b_{2 ,2},\cdots,b_{2,B} \}$, 
and $E'=
\cup_{i=1}^{B}(a_{1,i},t_1)\cup_{i=1}^{B-1}(a_{2,i},t_2)\cup\{(t_1,t_2)\}\cup_{i=1}^{B-1}(t_1,b_{1,i})\cup_{i=1}^{B}(t_2,b_{2,i})
$ (see Fig.~\ref{Fig:beta>0}).

Since $\beta$ is constant, this construction is done in polynomial time.

We have to show that there is a solution to the given instance $I$ of  3-partition iff there is a solution to the  problem  $GD(G^{\beta},B,m+2 \beta,\beta)$. 

If there is a solution to $I$ then by duplicating all  the $\beta$  edges ($t_1, t_2$) we get a partition $G^{\beta}$ into  exactly $m + 2 \beta$ subgraphs, each with exactly $B$ inputs.
Conversely, assume we are given a solution to $GD(G^{\beta},B,m+2\beta,\beta)$. 
The key point is that, according to Lemma \ref{lem:acceptable}, the only edges that can be duplicated in $G^{\beta}$ are the $\beta$ edges $(t_1,t_2)$ (see Fig.~\ref{Fig:beta>0}).
Moreover, since each of the dags $G'$ contains more than $B$ inputs , each of these  edges $x$ must be duplicated --  resulting in partitioning of $G'$ into two dags with exactly $B$ inputs in each part --  in order to get a solution that satisfies the constraints. Hence, all of these $\beta$ edges $(t_1,t_2)$ must be duplicated. Thus, we get $2\beta$ new subgraphs, and thus the remaining graph must have exactly $m$ parts, each with exactly $B$ inputs, which results  in a solution to the instance $I$ of the 3-partition problem. \section{Proof  of Theorem 5  }\label{appendix:theorem5sketch}
We use the same reduction as in the proof of Theorem 3, with a slight modification:
Each vertex satisfying $d_{in}(v) = d_{out}(v)$, is replaced by a subgraph.
For example, if $d_{in}(v) = d_{out}(v) = 5,$ then we replace it by the subgraph shown in Fig.~\ref{Fig:sketch5}.
The rest of the proof follows the same line.

\subsection{Proof of Theorem 6}\label{appendix:Theorem 6 sketch }
\begin{figure}[htbb]
    \centering
    \includegraphics[width=0.9\linewidth]{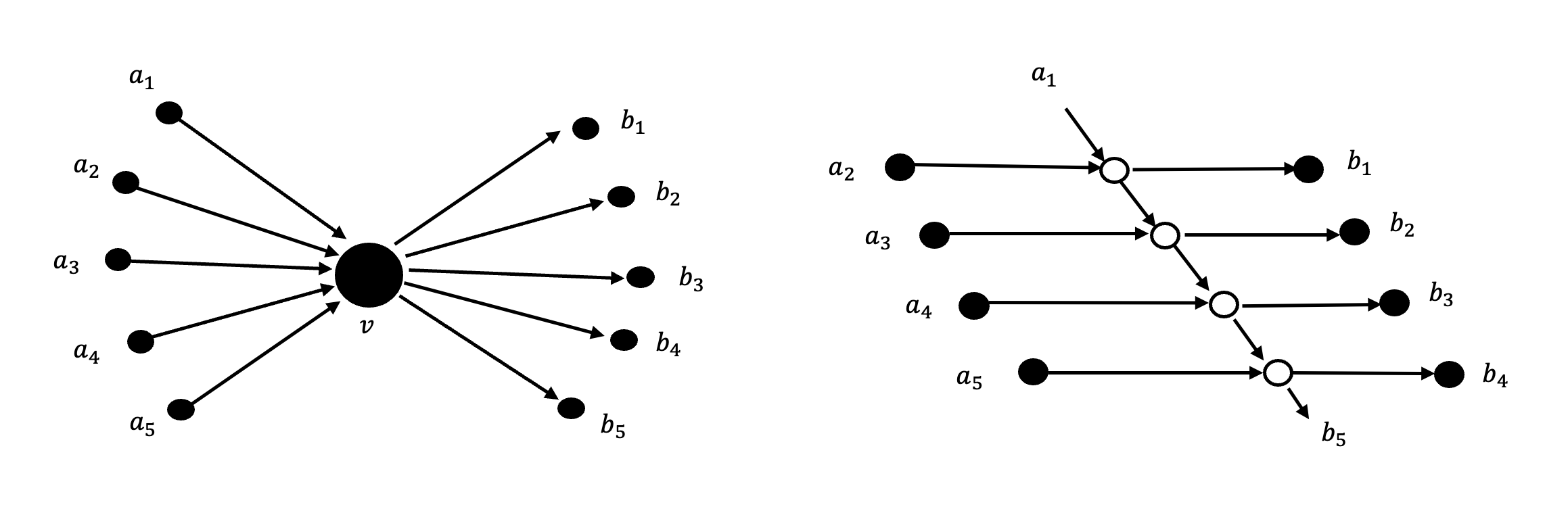}
    \caption{Illustration of the proof sketch for Theorem 5.}
    \label{Fig:sketch5}
\end{figure}

This is an extension of the proof of Theorem 4, where for each internal vertex  $v$ with  $d_{in} (v) = d_{out}(v)>2$ is replaced by the subgraph depicted in Fig.~\ref{Fig:sketch5}. The rest of the details follow the same line.

\section{The 3-partition NP-complete problem}\label{Appendix:3partition}

\noindent



\vspace{5mm}
\noindent
{\bf 3-partition.}
For our discussion, we refer to the book \cite{garey2002computers}.
\noindent We first use the following problem, known to be NP-complete \cite{garey1975complexity}:
\begin{problem}[3-partition problem]\label{prob:npc}
\par\noindent
\\$Input$:
A set $A$ of  $3m$ positive integers and a positive integer bound $B$, such that $\frac{B}{4} < a < \frac{B}{2}$ for every $a \in A$, and such that $\sum_{a \in A} a= mB$.

\noindent $Question$:
Can $A$ be partitioned into $m$ disjoint sets $A_1, A_2, ..., A_m$ such that $\sum_{a \in A_i} a=B$ for every $1 \leq i \leq m$ ? (note that each $A_i$ must therefore contain exactly three elements from $A$). 
\end{problem}

\end{document}